# A Techno-Economic Analysis of the Interconnectedness between Energy Resources, Climate Change, and Sustainable Development

MohammadReza Askari, Member, IEEE, Navid Parsa, Senior Member, IEEE

Authors are with Electrical and Computer Engineering Department, Shiraz University of Technology

**Abstract:** The rising global temperatures caused by climate change significantly impact energy consumption and electricity generation. Fluctuating temperatures and frequent extreme weather events disrupt energy production and consumption patterns. Addressing these challenges has become a priority, prompting governments, industries, and societies to pursue sustainable development and embrace eco-friendly economies. This strategy aims to decouple economic growth from environmental harm, ensuring a sustainable future for generations. Understanding the link between climate change, energy resources, and sustainable development is crucial. Techno-economic analysis provides a framework for evaluating energy-related projects and policies, guiding decision-makers toward sustainable solutions. A case study highlights the interaction between hydroponic unit energy needs, electricity pricing from wind farms, and product sales prices. Findings suggest that smaller 2-megawatt investments are more efficient and adaptable than larger 18-megawatt projects, proving economically viable and technologically flexible. However, such investments must also consider their social and environmental impacts on local communities. Sustainable development seeks to ensure that progress benefits all stakeholders while protecting the environment. Achieving this requires collaboration among governments, businesses, researchers, and individuals. By fostering innovation, adopting eco-friendly practices, and creating supportive policies, society can transition to a green economy, mitigating climate change and promoting a sustainable, resilient future.

1. Introduction

In recent years, the urgency to combat global warming and its devastating impacts has gained significant momentum on a worldwide scale. The international community is increasingly recognizing the necessity for collective action to tackle the ecological challenges posed by increasing temperatures, extreme weather events, and other consequences of climate change. In response, concerted efforts are being made to explore sustainable strategies that can effectively minimize these adverse effects and pave the way for a more resilient and sustainable future [1-5].

Within this context, the role of efficient energy utilization, renewable energy sources, and sustainable development has come to the forefront as pivotal elements in mitigating the environmental consequences of climate change. Embracing a sustainable approach to energy generation and consumption has become crucial in reducing greenhouse gas emissions and curbing the depletion of finite resources [6-8]. The transition towards renewable sources, such as solar, wind, and hydroelectric power, not only offers cleaner alternatives but also promotes energy security and independence. Sustainable development practices encompass a holistic approach that considers social, economic, and environmental aspects [9]. The concept of the eco-friendly economy, which lies at the heart of sustainable development, promotes the integration of environmentally friendly practices with economic growth [10-12]. This integrated approach seeks to decouple economic progress from environmental degradation, ensuring that economic prosperity aligns harmoniously with ecological well-being [13-15].

As scholars and policymakers delve deeper into understanding the intricate dynamics of climate change and its ramifications, multidirectional relationships between various factors have become a focal point of research [16-20]. The literature extensively explores the interconnectedness between global warming, energy resource utilization, sustainable development (including the eco-friendly economy), and the adverse environmental effects stemming from human activities. This inquiry aims to identify key factors and interdependencies, providing valuable insights for formulating effective strategies and policies to tackle climate change and achieve sustainability goals [21].

In alignment with this research pursuit, the objective of our work was to conduct a comprehensive investigation into the intricate network of multidirectional relationships highlighted in the article [22-25]. Through this endeavor, we aimed to make meaningful contributions to the existing body of literature in this domain. Our focus rested on two primary aspects [26]:

1) analyzing the nature of the multidirectional relationships among global warming, energy resources, sustainable development, and the eco-friendly economy, and proposing pertinent research areas for further exploration [27-30], and

2) applying a techno-economic analysis approach through a case study to provide concrete demonstrations of specific multidirectional relationships [31].

The article's subsequent sections were thoughtfully organized to delve into the core aspects of this investigation. The first section took an in-depth exploration into the correlation between global warming and energy consumption. By shedding light on the interplay between these two crucial elements, we aimed to uncover the implications of energy usage on climate change and vice versa [32-35]. Understanding this complex interrelation is essential in developing informed strategies to reduce carbon footprints and promote energy efficiency [36].

Moving forward, the second section delved into the intricacies of sustainable development, presenting an enlightening discussion on its broader significance within the context of multidirectional relationships. The exploration of sustainable development's multidimensional aspects allowed us to recognize its role as a catalyst for positive change in the face of global warming challenges [37-40].

The third section of the article presented a comprehensive coverage of the application of techno-economic analysis within the framework of a case study [38]. This analysis effectively showcased particular multidirectional connections, providing empirical evidence of the intricate web of relationships between global warming, energy resources, and sustainable development. The case study illustrated how the techno-economic approach can inform decision-making processes, facilitating the development and implementation of environmentally sound policies and projects [39-41].

Lastly, the concluding segment offered an all-encompassing discussion that encapsulated the principal discoveries and insights derived from the investigation. Besides summarizing the primary outcomes of the research, the concluding section also presented well-informed policy recommendations. These suggestions aimed to steer sustainable actions and decision-making at various levels, empowering governments, businesses, researchers, and individuals to play a role in building a more sustainable and resilient future [42-45]. The endeavors to combat global warming and foster sustainable development necessitate a holistic comprehension of the multifaceted relationships among key elements in the environmental and energy domains. Through this research, we endeavored to contribute to the worldwide pursuit of discovering viable solutions to global warming and nurturing sustainable development, with the aspiration of forging a more sustainable and prosperous world for the generations to follow [46].

## 2. Mutual connections between climate change and utilization of energy resources

The impact of human activities on the climate system is becoming more apparent, as recent emissions of greenhouse gases have reached unparalleled levels. This surge in emissions has led to a continual increase in global greenhouse gas (GHG) emissions, prompting the Organization for Economic Cooperation and Development (OECD) to predict a further 50% rise in GHG emissions by 2050. The primary driver behind this projection is the alarming 70% surge in energy consumption, primarily driven by the combustion of fossil fuels. Consequently, the pressing need to address the environmental harm caused by these activities has fostered a growing emphasis on exploring and harnessing alternative energy sources [47-50].

At present, fossil fuels hold a dominant role in meeting over 80% of the world's primary energy needs, making them a significant contributor to the excessive release of $CO_2$ into the atmosphere. This combustion process also results in the emission of other potent greenhouse gases like methane ($CH_4$) and nitrous oxide ($N_2O$), with methane being primarily produced during oil and gas extraction, accounting for nearly 10% of emissions from the energy sector. Additionally, nitrous oxide emissions are linked to various aspects of energy transformation, industry, transportation, and buildings, further exacerbating the global climate crisis [51-53].

Surprisingly, approximately one-fourth of the global population, roughly 1.6 billion people, still lack access to commercial energy sources, and countries such as China and India are experiencing exponential growth in energy consumption. This escalating energy demand calls for a shift towards sustainable energy consumption to mitigate the adverse effects of climate change. The significance of electricity production and consumption cannot be underestimated, as it significantly contributes to the utilization of energy resources, supporting national development and ensuring a certain level of well-being. With urbanization projected to significantly increase, it is estimated that 70% of the world's population will reside in cities by 2050, further emphasizing the urgency of addressing climate change [54].

To address the urgent issue of climate change, the International Energy Agency (IEA) has presented a bridging scenario consisting of five key strategies to reduce global energy-related greenhouse gas emissions. These strategies include improving energy efficiency in various sectors, such as industry, construction, and transport, and phasing out inefficient coal power plants while preventing the establishment of new ones. Another critical aspect of this scenario is the emphasis on investing in renewable energy technologies and ending fossil fuel subsidies for consumers. Additionally, concerted efforts are being made to reduce methane emissions from oil and gas production [55]. The IEA's analysis shows promising progress, with significant decreases in CO2 intensity in electricity generation projected for countries like China, the United States, India, the European Union, and Japan by 2025 compared to 2015. However, without changes to current industrial and energy-related policies, carbon emissions could lead to significant atmospheric temperature increases, reaching critical levels. To prevent such a dire scenario, it is crucial to stabilize carbon emissions at 450 parts per million to avoid a 2-degree Celsius temperature increase. Failure to do so could result in atmospheric intensity reaching 750 parts per million by 2050 [56].

Recognizing the paramount importance of preserving the future of humanity and the global environment, the sustainable utilization of renewable energy sources has become imperative. Renewable energy, including hydroelectricity, solar, biomass, wind, and geothermal power, provides a low-carbon solution to address climate change and foster a greener world. Encouraging extensive adoption of renewable energy requires implementing policies and regulations that promote its use. Many countries, both developed and developing, have already taken positive steps in this direction, introducing measures such as feed-in tariffs, renewable portfolio standards, subsidies, financial assistance, investment tax credits, and green certificates [57].

Additionally, monitoring energy usage and promoting conservation efforts can play a pivotal role in raising awareness among households about climate changes and energy-saving practices. By encouraging proactive involvement in combating climate change, countries can effectively reduce the hazardous effects of this global challenge. Following the guidelines set forth by the European Union, countries can reevaluate their energy policies, placing greater emphasis on sustainable energy utilization, renewable energy sources, and sustainable development. This collective effort will significantly contribute to minimizing adverse environmental impacts and ensuring a sustainable planet for future generations. The indisputable impact of human activities on the climate system has created an urgent need for decisive action in adopting sustainable energy practices, embracing renewable energy sources, and advancing sustainable development. By enacting green energy policies, fostering widespread adoption of renewable energy, and promoting energy conservation, we can collaboratively address climate change on a global level. These concerted endeavors are paramount in safeguarding our planet for current and future generations, cultivating a more sustainable and resilient world.

3. **China's policy-making towards the interconnectedness between energy resources, climate change, and sustainable development**

China's strategy formulation towards the interconnection between energy resources, climate change, and sustainable development stands as a pivotal and extensive undertaking, profoundly impacting the nation's trajectory and contributing significantly to global environmental well-being. Confronted with the urgent necessity to address the formidable challenges presented by climate change and the exhaustion of finite resources, China has embraced a determined green growth strategy, charting a course towards a more sustainable future anchored in renewable energy sources.

At the core of China's policy endeavors lies a resolute ambition to diminish its reliance on fossil fuels and harness the potential of renewable energy. Ambitious objectives have been established to propel the implementation of renewable energy technologies, aiming to secure a substantial portion of electricity generation from wind, solar, and hydro power. With this strategic emphasis, China envisions a profound decrease in greenhouse gas emissions, curbing its impact on the global climate crisis and embracing a leading role in combatting climate change.

The dedication to sustainability and energy transformation is echoed in the manifold measures implemented by China. Energy efficiency and conservation have become fundamental principles of this progressive policy approach. Through rigorous energy-saving regulations for industries and visionary building codes prioritizing energy-efficient construction, China is cultivating a culture of prudent resource utilization. Citizen awareness is heightened through campaigns promoting energy conservation, urging individuals to embrace eco-conscious behaviors in their everyday lives.

To reinforce its endeavors towards sustainability, China has made substantial investments in research and development, igniting a fervent pursuit of cutting-edge clean technologies. The nation's dynamic innovation ecosystem seeks to unlock revolutionary solutions for sustainable energy production and consumption. Embracing collaboration and open markets, China actively invites private sector participation and foreign investment to nurture a robust and vibrant green technology landscape.

China's unwavering commitment to sustainable development transcends national borders, embracing a global perspective. Actively engaging in international climate negotiations, China contributes its voice and resources to worldwide efforts to combat climate change. As a foundation of its global vision, China's Belt and Road Initiative champions the cause of sustainable energy infrastructure in partner countries, fostering a network of interconnected energy systems grounded in environmental responsibility.

China's forward-looking strategy formulation towards the interdependence of energy resources, climate change, and sustainable development is an exemplary embodiment of its dedication to constructing a greener and more sustainable future. By steadfastly prioritizing the deployment of renewable energy, optimizing energy efficiency, and fostering international cooperation, China emerges as a formidable force in the ongoing global transition towards a low-carbon and sustainable energy future. With unwavering determination, China endeavors to leave a lasting legacy of environmental stewardship and prosperity for the present and future generations, inspiring a world where sustainability forms the bedrock of progress and shared prosperity.

### 4. The concept of sustainable development and its interconnected aspects

This section explores the concept of sustainable development from an interdisciplinary perspective, focusing on environmental economics, green growth, and considering social and political dimensions such as democracy, politics, and decision-making. Environmental resources can be broadly categorized into living (biotic) and non-living (abiotic) resources. Living resources include ecosystems, agricultural products, forestry, fisheries, plants, animals, and microorganisms, while non-living resources encompass water, minerals, sunlight, wind, and other renewable natural resources like solar energy, as well as recycled resources and non-renewable, non-recyclable resources such as fossil fuels. Scarcity of resources like food, water, energy, land, and rare materials can lead to competition among consumers. Human activities have the potential to alter environmental conditions, affecting factors such as soil acidity, nutrient content of surface waters, radiation balance, and concentrations of rare elements in the atmosphere and food chain. To address the critical loss of biodiversity, climate change, and other adverse environmental impacts, it is essential to adopt a more sustainable approach to food production, transportation, and housing. Urgent mitigation measures are required to combat pressing environmental issues like habitat diversity, excessive use of renewable resources, climate change, and airborne particulate pollution. Undesirable environmental effects often arise as unintended consequences of economic activities. The projected doubling of the world's population by the mid-21st century is expected to significantly increase global demand for energy services, posing a significant challenge to economic sustainability. At the same time, over one billion people still lack access to electricity, presenting a major global challenge in ensuring modern energy services. Addressing persistent energy disparities is crucial for effective environmental conservation in such contexts. To avoid detrimental competition for resources, it is vital to prioritize effective management of undesirable environmental impacts. By promoting sustainable practices, embracing green growth strategies, and ensuring inclusive access to modern energy services, we can work towards a more harmonious and sustainable coexistence with our environment.

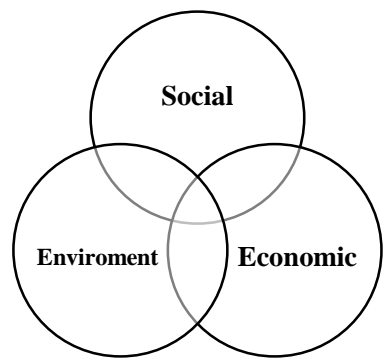

**Fig. 1:** Interplay of the environmental, economic, and social aspects of sustainable development

### 4.1. The process of sustainable development in relation to relevant concepts

Sustainability involves a well-balanced approach to economic, societal, and ecological well-being. To establish a sustainable world, it is crucial to responsibly utilize ecological resources like materials, energy sources, land, and water at a pace that allows for their replenishment. Economic viability is accomplished when businesses or nations efficiently and ethically utilize their resources, enabling continuous operations and sustainable profits. Societal viability is achieved when all communities and societal structures achieve a high level of well-being. While sustainable progress and continuity are closely connected, sustainable progress can be viewed as a pathway towards achieving overall continuity. The researchers conducted a study analyzing the distinctions between "continuity" and "sustainable progress" in the context of media discourse in Germany. The research aimed to comprehend how the media defines continuity and sustainable progress in the German context. Sustainable progress, as a distinct research field, explores the dynamic interactions between ecological and societal systems, contributing to a deeper understanding of these relationships. It also encompasses specific approaches to transforming ecological and societal systems towards greater continuity. Despite the widespread use of the term sustainable progress in policy-making over the past 25 years, ongoing debates persist about its precise meaning and practical implications. The terms continuity and sustainable progress are inherently complex and ambiguous in their meanings. The article delves into three research domains investigating how the media, particularly newspapers, address continuity topics.

- Firstly, some research focuses on the broader field of sustainability science and sustainable communication without explicitly using the term "sustainability." Instead, it concentrates on specific aspects within the broader context of sustainable development, such as climate change and biodiversity loss in ecosystems.
- Secondly, other research emphasizes presenting various indicators of sustainable development in the media.
- Finally, the article specifically examines how the term "sustainability" is used and understood in media discourse.

The notion of durability stems from ecology, pertaining to an ecosystem's capacity to endure over time. Sustainable progress, commonly defined, seeks to fulfill current requirements without compromising the ability of future generations to satisfy their own requirements (International Institute for Sustainable Progress - IISD). Achieving this necessitates collaborative efforts, such as the Paris Agreement, to offer financial aid and technology exchange mechanisms, empowering developing nations to decrease greenhouse gas emissions and restrict the global temperature rise to well below 2 degrees Celsius above pre-industrial levels. Another research [18] introduces a novel conceptual framework through a multidisciplinary analysis of sustainable progress, critiquing ambiguous definitions. It links durability to ecology and the Brundtland Report [19], also known as our mutual future report, which functions as an environmental "logo." In the Brundtland Report [19], development is linked to the economy, portraying an economic "logo" [18]. It also explores the concept of natural capital, defined by Pierce and Turner [20] as all environmental assets and resources, ranging from oil reserves in the ground to soil and groundwater quality, from oceanic fish reserves to the Earth's capacity for carbon sequestration. Hence, natural capital encompasses non-renewable resources, the limited potential of natural systems for renewable resource production, and the absorption of emissions and pollutants [18]. The concept of unchanging natural capital, known as "robust sustainability," becomes vital in providing a criterion for sustainability. It [18] also emphasizes the significance of fairness, as highlighted [21], stating that achieving long-term economic or environmental sustainability becomes extremely challenging in the absence of social justice. Achieving sustainability involves striking a balance between social, environmental, and economic objectives. The literature also considers intergenerational and intragenerational justice, ensuring fair distribution of resources between current and future generations, establishing intergenerational equity. It examines several studies that integrate social, economic, and environmental concerns in sustainable development planning and management [18]. Contrary to the prevailing perspective that economic goals (like poverty alleviation and economic growth) should take precedence over environmental objectives, our common future [19] insists that environmental health serves as a prerequisite for social and economic prosperity. It [24] raises concerns about the term "sustainability," pointing out its ambiguity regarding whether it refers to general development, green economic growth, or human well-being. Sustainability indicators are assessed within the framework of ecological sustainability, economic sustainability, and sustainable development. However, there remains no clear definition of ecological sustainability and sustainable development. On the other hand, economic sustainability refers to steady-state consumption [24], leading to confusion in distinguishing economic well-being from human well-being and differentiating between green growth resources and sustainable development. It suggests that emphasizing green growth holds greater significance than other abstract concepts such as sustainable development or universal wealth due to its measurability. Similarly, Hopwood et al. [25] also discuss the need for a more precise definition of sustainable development. While sustainable development holds various meanings, it is commonly understood as "an attempt to integrate increasing environmental concerns with social and economic issues across a wide spectrum" [25]. This contradicts the prevailing view that environmental and social-economic issues should be treated separately. Hopwood and colleagues [25] argue that sustainable development gained prominence because it drew attention to environmental issues, social-economic concerns such as poverty and inequality, and the aspiration for a healthier future. Consequently, they emphasized sustainable development. Despite claims that sustainable development is an essential tool for global integration, the ambiguous definition of sustainable development in the Brundtland report leads businesses and governments to prioritize economic growth for poverty eradication. Another research [26] disagrees with this perspective, as it cannot be sustained in a world with limited ecosystems; thus, achieving qualitative development requires moving away from the quantitative approach presented in the Brundtland report. It [27] analyzes the social aspect of sustainable development in relation to its links with the environmental component. To accomplish this, it proposes a conceptual framework that defines four social concepts: public awareness, equality, participation, and social coherence. As the link between social and environmental aspects of sustainable development remains underdeveloped, extending the parameters of the social aspect and connecting it to the environmental requirements becomes necessary. Justice also holds a significant place in the sustainable development literature, as it pertains to the equitable distribution of welfare resources. Justice in terms of life chances is also crucial, indicating that all citizens should have equal opportunities for survival and the realization of their development potential regardless of gender, ensuring fair redistribution The comprehensive notion of equity encompasses ensuring access to pure water, sustenance, employment, education, housing, necessary medicines, unpolluted surroundings, social network availability, and promoting freedom from discrimination based on gender, religion, or race, among other factors. Another critical idea in the sustainable development literature is sustainability awareness. Policies to raise awareness play a crucial role in encouraging individuals to adopt sustainable consumption patterns instead of conventional ones.

Instances of such initiatives include eco-friendly advertising campaigns, environmental labeling, awareness-raising events, environmental education programs, and sustainability education programs. The underlying belief is that consumers will actively seek opportunities to adopt more eco-conscious lifestyles and household practices. Participation emerges as a crucial concept in the realm of sustainable development literature. It emphasizes the importance of involving various social groups in decision-making processes. By integrating these groups into participatory endeavors, increased public engagement can lead to social coherence and social sustainability. Social coherence holds a significant role in the discourse of social policy, encompassing aspects such as heightened happiness and well-being, reduced social conflicts and crime rates, enhanced interpersonal trust, and suicide prevention within society.

While social harmony is connected to various aspects of social advancement, its correlation with environmental demands is relatively weak. Nevertheless, policies and initiatives aimed at achieving sustainable development underscore the connection between social cohesion and environmental goals. Conversely, environmental factors can have adverse effects on social cohesion. The article proposes that by strengthening these social foundations, acceptance of environmental, international, intergenerational aspects, and sustainable development indicators can be promoted. However, it is crucial not to disregard the institutional aspect, as social sustainable development can be effectively overseen through institutions and/or organizations [2].

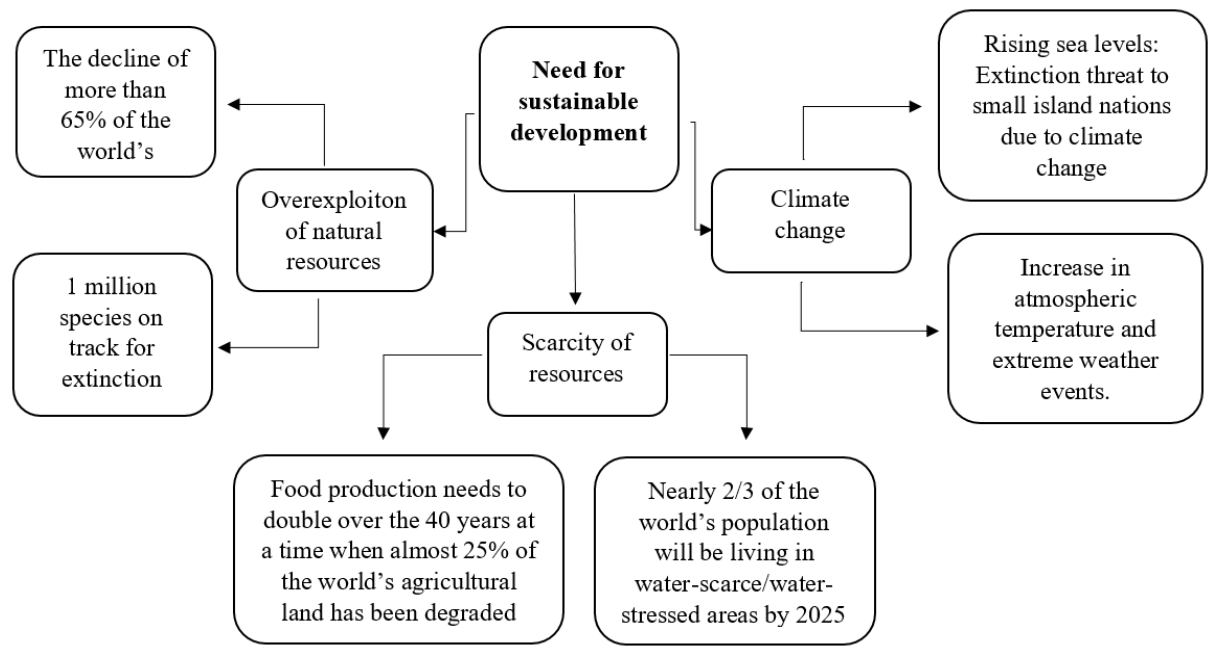

**Fig .1:** Need for sustainable development

### 4.2. An overview of the green economy

The green economy carries great importance and should be supported by government policies and regulations to break the harmful cycle of climate change. Its main aim is to enhance social well-being and equality while addressing environmental risks and constraints that hinder ecological growth. It emphasizes social responsibility and efficient use of natural resources, leading to lower carbon emissions and the adoption of clean energy solutions. Prioritizing environmental values and sustainable energy solutions is crucial for achieving productivity and meeting the demands of the century.

The concept of the green economy was initially explored in a research [29], and its usage has become more widespread in recent times. In 1992, it was defined as positioned between deep ecological extremism and pure neoliberal thinking. According to the United Nations Environmental Program [20], the green economy seeks to promote economic growth while preserving the environment. Decoupling is a key idea in this context, emphasizing the need to separate economic growth from excessive resource consumption. Energy production plays a vital role in achieving a green lifestyle and economy, necessitating the use of natural resources for energy generation without harming essential life processes. Hence, the relationship between economic activities and environmental impacts must be carefully managed, with a focus on sustainable resource use. The Organization for Economic Co-operation and Development (OECD) [21] suggests that decoupling requires breaking the link between economic growth and its negative environmental effects.

As described by the United Nations Environmental Program (UNEP) [22], decoupling involves reducing resource usage during economic growth and disentangling environmental degradation from economic development. It aims to separate environmental impacts from economic activities and energy utilization. Decoupling can be either absolute or relative. Absolute decoupling is achieved when economic growth is accompanied by a decrease in resource usage, while relative decoupling occurs when the increase in resource usage is slower than the rate of economic growth [22]. While some countries have achieved relative decoupling, absolute decoupling remains relatively rare.

Another study [23] explores the sustainability of economic growth and convergence in regions at different stages of development. The convergence process involves two aspects:

- First, it addresses the lagging position of poorer countries and regions compared to their wealthier counterparts.
- Second, it explores the possibility of fostering more sustainable patterns of economic growth in developing countries.

The latter aspect is of great importance as it presents opportunities for using fewer resources, resulting in reduced greenhouse gas emissions and pollution. The article also explores the implications of the convergence process, which have both economic and environmental consequences.

Examining the economic and environmental progress of developing countries reveals a correlation between economic setbacks and increased energy consumption and $CO_2$ emissions. It is crucial to investigate whether less affluent countries or regions are catching up with wealthier nations, but equally important to consider whether they are following inefficient traditional development paths that impose significant environmental pressures. Thus, giving special attention to the environmental aspects and consequences of convergence becomes imperative.

The article underscores the high significance of sustainable growth for the economies of developing countries. Sustainable growth can address economic disparities, political instability, and persistent conflicts. However, the environmental consequences of growth and convergence vary. Developed countries display decoupling of economic growth from significant environmental effects, particularly through absolute decoupling of $CO_2$ emissions. It is crucial not to overlook the possible transfer of environmental pressures from developed to developing countries and to ensure that positive decoupling trends are established in the latter.

In most developing countries, relative decoupling of environmental impacts from economic growth is evident. However, this is insufficient, as their environmental impacts increase at a much faster rate compared to those of developed countries. This indicates that developing countries or regions may eventually reach pollution levels and resource-intensive practices similar to those seen in developed countries, highlighting the need for investment in green initiatives to alleviate pressures on natural resources in the developing world.

Transitioning to a green economy requires adherence to technical standards and the implementation of subsidies, while also monitoring organizational challenges and the outputs of green economic practices. These measures can either discourage or encourage green activities within each institution. Additionally, a research [18] introduces the concept of "Ecoform" as a desirable ecologically-oriented form for human habitats. Ecological design and the definition of urban structures to foster environmentally friendly and sustainable environments are of great significance. It also refers to sustainable development literature that discusses various technologies and ideas related to the environment and sustainability, including alternative building materials, renewable energy, organic foods, conservation, and recycling. Many researchers argue that achieving an ecological approach through the design of buildings, communities, cities, and regions necessitates considering energy efficiency. Such designs have the potential to mitigate air pollution and enhance energy productivity. A concrete example of the connection between renewable energy, organic food production, and climate change is presented within this context.

Green energy plays a prominent role in sustainable development literature, as it can fulfill energy requirements for both industrial and local purposes. Therefore, the development and implementation of sustainable green energy strategies and technologies are of utmost importance. If green energy resources and technologies are widely adopted by developed and developing countries, it is likely to lead to sustainability. Green energy resources and technologies are vital for sustainable development due to their lower environmental impact compared to other energy sources. Furthermore, they are non-depletable, and their decentralized nature makes local solutions advantageous for utilizing green energy resources.

Consequently, investment in green energy and related technologies, along with the implementation of sustainable green energy strategies, holds significant significance. Various strategies can facilitate industrial and technical progress during the transition to green energy technologies. These strategies include analyzing green energy resources, providing governmental and public support for the green energy economy, managing the production, consumption, distribution, transformation, and marketing of green energy, conducting research and development, utilizing sustainable green energy technologies, and designing and executing practical ecological and environmental programs centered around green energy.

Highlighting the relationship between a green economy and sustainable development is crucial. Bina [4] explores the meaning and consequences of a green economy for sustainable development and argues that greening the economy and promoting growth are essential for improving the global economic condition and preventing crises. The terms "green economy" and "green growth" are used interchangeably to signify low-carbon development and an eco-industrial framework that preserves ecosystem well-being. The United Nations Commission on Sustainable Development (UNCSD) [5] issues a warning that a green economy combined with sustainable development poses a significant challenge for the present world, requiring more thoughtful consideration. For instance, the strategies adopted at the Rio+20 Summit for green economies and sustainable development were found to be incongruous.

The Rio+20 Summit recognized the green economy as a crucial tool for achieving sustainable development, eradicating poverty, promoting social inclusion, improving human welfare, creating job opportunities, and conserving the reliable functioning of Earth's ecosystems. According to it [20], sustainable green growth can be achieved by raising environmental awareness among producers and consumers. Green growth strategies can address the needs for food, transportation, housing, energy, and water in economies and communities. They also mitigate adverse shocks related to resource consumption and environmental impacts. Environmental innovations stemming from renewable alternative technologies enable industrial production without compromising environmental preservation.

Research and development are particularly pivotal in achieving these objectives and necessitate budgetary support.

## 5. Cognitive approach to multivariate analysis

In the pursuit of sustainable development, a wide array of notions and concepts related to eco-friendly energy has been proposed globally. These notions encompass the replacement of coal/oil-based facilities with solar, wind, and hydropower farms. However, putting these ideas into practice is not always straightforward. Some countries are making noteworthy efforts to achieve their national targets of incorporating 65%, 45%, or even 100% renewable energy into their electricity generation mix, surpassing significant milestones in reducing fossil fuel dependency.

The overall vision of achieving a mix of costless energy by 2050 or 2060, spanning across heating, electricity, and transportation sectors, is both promising and ambitious. As a result of various hopeful and ambitious energy plans, there has been a remarkable increase in the installation of wind farms and solar parks. Nevertheless, the immediate pursuit of interconnection capacities with neighboring countries has not been prioritized, leading to congested electricity grids and simultaneous generation, resulting in instances of curtailment, negative pricing, and energy wastage.

As the share of wind and solar energy continues to rise, there will be ongoing efforts to reduce and eliminate regulatory barriers. In practical terms, during the past few years, investors have faced either minimal profits or even losses. Consequently, renewable energy investments are now under global pressure, prompting investors and stakeholders to seek immediate and viable solutions. Achieving a genuine and all-encompassing transformation requires more than just an eco-friendly energy system. It necessitates connecting energy resources with sustainable development and their influence on climate change. To achieve this, laying the foundation and conducting a comprehensive analysis based on the integration of 100% renewable energy sources (RES) are essential, along with proposing a sustainable resolution to the mentioned issue.

By implementing a demand-responsive energy-food-transportation nexus, it becomes feasible to incorporate substantial surplus capacity while simultaneously generating locally sourced products with minimal $CO_2$ emissions. The excess energy that needs to be controlled, as indicated by system operators, can be stored as hydrogen in trucks and utilized as hydrogen fuel for transportation, or it can be supplied to factories for decentralized vegetable production. The system can supply the required power to support power plants during peak-demand periods when electricity prices are high, and during low-price periods, it can store electricity for future use.

The widespread deployment of such systems and active participation in demand response (DR) programs can significantly impact the seamless integration of RES into intelligent regions, fostering sustainable development.

## 5.1. A case study in China

Extensive expansion within such frameworks can guarantee a uniform and sustainable yearly electricity generation, irrespective of weather conditions. These systems can function as virtual storage power plants, advocating the use of renewable energies and minimizing $CO_2$ emissions. This investigation concentrated on evaluating the requirements of hydroponic units, product selling prices, wind farm electricity costs, and concentrated energy demand.

After the wind resource analysis, the ultimate design of wind farms can be determined, and a financial assessment can demonstrate the benefits for stakeholders. To identify locations with high wind speeds near a specified city in northern China, accessible and appealing to investors, various factors like ecological constraints and spatial requirements (e.g., streets, lakes, rivers) were considered, necessitating the utilization of Geographic Information Systems (GIS) tools.

To assess the wind conditions in the designated city, WindRose and WAsP (Wind Atlas Analysis and Application Program) were utilized. Based on statistical analysis of the broader region using annual wind measurements from a monitored mast, the maximum average wind speed was approximated to be about 8.34 meters per second at 98 meters above ground level (7.38 meters per second at 7.87 meters above ground level). The minimum wind speed in the wider area was 3.4 meters per second. Wind measurements were collected for a year, adhering to international standards, using a 30-meter mast. All information was stored in a data logger and transmitted to a certified laboratory through a common identification module (SIM card). The annual wind speed uncertainty, based on available data, was 1.11 meters per second, with a recorded maximum gust of 50.5 meters per second and a maximum 10-minute average wind speed of 45.4 meters per second.

WindFarm and WAsP tools were employed to generate precise wind resource maps of the broader region. Figure 4 presents the estimated wind speed image, indicating various locations suggested for wind turbines (also marked on the elevation map).

Numerous sites with exploitable wind speeds (e.g., wind speed > 7.5 meters per second) were identified. Areas colored light blue, blue, and green exhibited wind speeds lower than 6.5 meters per second, while areas colored yellow, orange, and red fell within the range of 7.5 to 8.34 meters per second, capturing the attention of potential investors. Profitable wind energy projects were those that contributed to increased wind energy integration and environmental protection. The analysis identified the available regions for wind turbines and determined the potential for large-scale hydroponics and energy storage in the broader area. However, the impact would be restricted, with delivery options mainly suitable for short distances, catering to the specified city (located less than 8 kilometers away) due to harsh winters causing shortages of fresh vegetables and leafy greens in the region.

China had nearly 3.7 gigawatts of installed wind energy, with a target of 8.5 gigawatts. Consequently, several new projects were scheduled for the upcoming years, leading to a reduction in the wind energy reduction rate. Germany and the UK experienced an increased wind energy reduction rate (6-7% and 7-8% respectively, including 17-18% for onshore Scottish wind farms). According to the transition to green energy trading for all Europeans, wind farms were expected to avoid fines for reducing their annual energy production (AEP) if grid balancing complied with market regulations. Therefore, in normal conditions, all limited power would be utilized. However, this concept assumed an average 5% reduction applied to existing apartments, operating hydroponic units at lower electricity prices (0.05 and 0.07 EUR/kWh). The cash flow analysis estimated that 85.5% or 85.0% of the AEP would be delivered to the electricity market, with 5% or 0% allocated to hydroponic units in various apartments in the specified city.

Based on the analyses, it was observed that although total income was higher in higher-capacity wind farms, the internal rate of return (IRR) was significantly more profitable in lower capacities (e.g., 5.2 MW). It is essential to highlight that with an 28 MW project, 1968 apartments could benefit from the reduction. Assuming one worker was required for every 30 apartments, this would result in 105 people being employed through a concentrated hydroponics venture in the city. Opting for a 3 MW project would only employ 20 people. Hence, in general, the first option could be considered a more sustainable solution for the local community, establishing a deeper connection between energy resources and sustainable development.

Another critical outcome was the reduction of $CO_2$ emissions. In general, the daily $CO_2$ emissions average in hydroponic units was significantly lower compared to greenhouses. This mass deployment plan would significantly reduce $CO_2$ emissions from transportation since long-distance transportation of food within the urban community would be unnecessary, as it could be sourced from nearby locations within the city. This relationship between energy resources and climate change was inherently evident in this regard. Below is a summarized table of the key points from the 4.1. text:

**Table. 1:** The table provides a concise overview of the main points in the text. Some details may have been omitted for brevity.

| Topic | Summary |
| --- | --- |
| **Objective** | Study focuses on ensuring sustainable electricity generation, virtual storage power plants, and minimizing $CO_2$ emissions. |
| **Wind Resource Analysis** | WindRose and WAsP tools used to evaluate wind speeds near specified city in northern China. Estimated wind speed: 3.4 to 8.34 m/s. |
| **Wind Farm Design** | Final design of wind farms determined based on wind resource analysis. |
| **Potential Wind Farm Locations** | Identified various locations with exploitable wind speeds (>7.5 m/s) that attract investors. |
| **Wind Energy Projects** | Profitable projects contribute to increased wind energy integration and environmental protection. |
| **Potential for Hydroponics** | Analysis identified potential for large-scale hydroponics and energy storage in the region. |
| **Delivery Options** | Delivery options mainly suitable for short distances, serving the specified city due to harsh winters causing food shortages. |

| **China's Wind Energy Growth** | China had 3.7 GW installed wind energy with a target of 8.5 GW. Several new projects planned for the future. |
|---|---|
| **Wind Energy Reduction Rate** | Germany and the UK experienced an increased wind energy reduction rate (6-7% and 7-8% respectively, including 17-18% for onshore Scottish wind farms). |
| **CO2 Emissions Reduction** | Hydroponic units significantly reduce daily CO2 emissions compared to greenhouses. Mass deployment plan reduces CO2 emissions from food transportation. |
| **Economic Considerations** | Higher-capacity wind farms generate more income, but lower capacities have higher internal rate of return (IRR). |
| **Employment Impact** | Larger hydroponic project (28 MW) can employ 105 people, while smaller project (3 MW) employs 20 people, offering a sustainable solution for the local community. |

6. **Conclusion**

In conclusion, the rapid exhaustion of the world's natural wealth reserves poses a serious threat to the well-being of future generations, creating significant challenges that must be addressed urgently. Through a thorough analysis of the intricate interconnections between global issues concerning climate change and energy resource utilization, it becomes apparent that a fundamental shift in international policies is necessary to foster a green growth strategy. This evaluation not only provides valuable guidance to a country's economic sectors but also offers specific insights for the largest and fastest-growing economies, which owe their success to political stability and good governance.

By embracing greener economies, this comprehensive and interdisciplinary analysis illuminates the current social and economic costs related to energy demand and supply, emphasizing the need for improved management practices. The overarching aim is to tackle a broad range of energy policy challenges confronting the world, including achieving the millennium development goals aimed at combating climate change.

While some may consider neglecting the well-established correlation between socio-economic development and environmental sustainability as a means to meet current needs, such a short-sighted approach risks compromising the ability of future generations to meet their essential requirements. The earlier analysis has highlighted that although a clear technical solution exists, a more optimal and straightforward alternative emerges when considering other critical factors like stability and climate impact, which should guide decision-makers towards making informed choices among the available options.

Despite the prevailing reliance on fossil fuels in energy production due to the slow pace of innovation and technological shifts, the implementation of new energy resources on an industrial revolution scale becomes imperative to address the urgent challenge of reducing CO2 greenhouse gas emissions linked to energy production before 2050. Detaching energy production from fossil fuels opens up exciting opportunities for new green industries, technological innovations, and transformative structural changes that are integral to transitioning to a sustainable green economy.

The insightful analysis, summary, and additional perspectives offered by the techno-economic analysis method demonstrated through the case study on wind farm projects showcase not only the viability of renewable energy solutions but also exemplify potential global or regional scenarios for sustainable development.

Furthermore, individual investments in hydroponics have already shown positive outcomes, leading to reductions in $CO_2$ emissions and electricity costs. The study provides valuable insights into meaningful supply and transportation solutions for food, an essential aspect of sustainable development.

This study plays a vital role in elucidating the intricate relationship between energy resources, climate change, and sustainable development, with a clear objective of achieving global energy and greenhouse gas emission reduction goals. It is important to note that the study's intention is not to provide definitive answers, but rather to maintain objectivity and provide a sound basis for policy-making. The multi-dimensional analysis approach adopted in this study is one among various investigation methods, encouraging further exploration and comparison of alternative policy pathways for a sustainable and resilient future. It is through such collaborative efforts and informed decision-making that we can build a world where economic prosperity, environmental preservation, and social well-being go hand in hand for the benefit of present and future generations.